\documentclass[review]{elsarticle}
\usepackage[hyphens]{url}
\usepackage{hyperref}
\hypersetup{breaklinks=true}

\begin{document}

\begin{frontmatter}

\title{Rapid loss of Arctic sea-ice}

\author{Beril Sirmacek}\ead{bsirmacek@gmail.com}




\begin{abstract}
Several environmental tipping points and self-reinforcing feedback loops are still disregarded within the frequently used climate models. Thus, existing climate models are not very representative for providing projections of the conditions after the actual environmental tipping points are triggered. In this contribution, we discuss the presence of the Arctic sea-ice loss tipping point and its influence on global warming. Our literature search and PIOMAS sea volume observation data based projection indicate the significant acceleration of the sea-ice loss (which leads to a year around and permanent ice-free Arctic) in near future.     
\end{abstract}


\end{frontmatter}


\section{Introduction}
The Arctic sea-ice is crucial for a variety of reasons. It keeps the climate balance, regulates the global temperature and circulates earth's both atmospheric and oceanic currents. As the Arctic is considered to be a barometer of global climatic change, a subtle change in the sea-ice becomes extremely crucial for the rest of the earth’s climate \cite{KHARE20211}. The recent decrease of ice coverage is one of the most striking indicators of global environmental change \cite{Wang2020}. 

Due to the phase change from the solid sea-ice to liquid state without temperature change (known as \textit{latent heat} or \textit{insensible heat}), in satellite based observations no significant temperature anomaly is visible during the summer months of the Arctic. However, lately higher anomalies are seen in autumn period \cite{Yadav2022}. This indicates that the Arctic sea-ice loss is accelerating, contrary to how the latest IPCC AR6 full report prediction models indicated \cite{IPCC2021}. We notice that the \textit{latent heat} component of the melted ice within the ocean is not used as a parameter of the sea-ice loss prediction models, which will be especially impactful after the first ice-free summer. The \textit{latent heat} content, which is now used to melt the ice, will be replaced by sensible heat to warm the ocean's water once the Arctic faces her first ice-free summer. The latest IPCC AR6 report has indicated that "only occasionally ice-free summers" could be observed in the Arctic at least once before 2050 under all assessed SSP scenarios. The AR6 report has also indicated that Arctic summer sea-ice varies approximately linearly with global surface temperature, implying that there is no tipping point and observed/projected losses are potentially reversible. This statement is labelled as "high confidence" within the IPCC AR6 report. We believe that this strong statement brought a false sense of trust on the reversibility of the Arctic sea-ice loss. Possibilities of reverting the sea-ice loss event, with reducing $CO^2$ emissions became a target and globally accepted by governments, industry and repeated by mainstream media to convince the world population. Even though we do agree that $CO^2$ emissions must be reduced, herein we show that the Arctic sea-ice loss is not reversible. This also proves that Arctic holds a major climate tipping point (being the major controller of the global climate system), contrary to the IPCC AR6 report (which claims that 'there is no tipping point in the Arctic'), Arctic appears to be a climate tipping point as well as being a part of the self-reinforcing feedback loops.

The remainder of the paper is organized as follows. The background of the study and the literature findings are summarized in Section 2. In Section 3, we provide our projection method using PIOMAS sea-ice volume data \cite{Schweiger2011}. Using the sea-ice volume data from 1979 to 2021, we analyze the \textit{latent heat} energy budget and sea-ice volume projections for the upcoming years. In Section 4, we discuss our results, we underline the existence of the climate tipping point in Arctic and we compare our conclusions with the IPCC AR6 report's related statements. Finally, in Section 5, we conclude our study by suggesting further action plans for the scientists and the policy makers.

\section{Background}

\subsection{Global warming and its relation to the Arctic sea-ice}

Arctic sea-ice moderates the temperature of the ocean and atmosphere. In an ice-free Arctic situation, the earth will not only experience rapid and global impact of changing atmospheric and oceanic cycles globally, but also the global average temperature would rise rapidly until the earth system reaches a new thermodynamic equilibrium \cite{Alley2000}. Such impacts have been observed in the past and it is happening much rapidly in our current times \cite{BrighamGrette18431}. Arctic sea-ice extent recorded a dipping minimum in 2020 (3.74 million km2), which is 0.35 million $km^2$ more than the lowest record of 2012 (3.39 million km2) \cite{NSIDC}.

\subsection{Self-reinforcing feedback loops and the past Dansgaard–Oeschger event}

Steffen et al. \cite{Steffen8252} explained the risk of self-reinforcing feedback loops that could push the Earth System toward a planetary threshold that, if crossed, could cause continued warming of the planet even as human emissions are reduced. Such self-reinforcing feedback loops earlier have been observed and some of them known as Dansgaard–Oeschger events (D–O events). These events are rapid climate fluctuations that occurred 25 times during the last glacial period. In the Northern Hemisphere, they took the form of rapid warming episodes, typically in a matter of decades, each followed by gradual cooling over a longer period. For example, about 11,500 years ago, average annual temperatures on the Greenland ice sheet increased by around 8C over 40 years, in three steps of five years, where a 5C change over 30–40 years was noticed \cite{Bond1999}. Our current global climate prediction models do not correctly take into account such self-reinforcing feedback loops, besides the significant effects of aerosols on clouds on Earth's overall energy balance \cite{Rosenfeld2019}. Besides seeing the likelihood of a new D-O event which might start occurring from higher temperature levels, the current earth system contains more anthropogenic content (like nuclear power centers \cite{KAUKER201613}) which might trigger damaging environmental events and global warming even further \cite{sirmacek2022}.

\subsection{Current climate models for predicting Arctic sea-ice loss and the accumulated ocean heat content}

Ruby Leung et al. \cite{Ruby2006} discussed the need of the existing climate models for accurate predictions which cannot be done with the existing models. They mentioned that 'upscaling' and 'downscaling' of the climate models must be studied before more accurate models can be developed. With 'upscaling' suggestion, they highlighted the need of finding the connection of the local drought, wildfire, volcano, tornado like events to global changes. With 'downscaling' suggestion, they highlighted the need of creating more precise models for certain local regions (just like Arctic sea-ice covered areas) to model their changes better. One major limitation of most climate models is that they don’t give a complete picture of the physics that governs interactions between sea-ice, the ocean and the atmosphere. One reason is that the processes are new areas of research and nobody has studied them in detail yet \cite{PIDCOCK2014}. Cohen et al. \cite{Cohen2020} expressed that the connections between Arctic Amplification (AA) and midlatitude temperatures are not understood. Due to a number of the missing interrelations and unknowns in the self-reinforcing feedback loops, the existing climate models provide sea-ice projections which show slowing down in decreasing path (instead of acceleration) and never reaching to zero ice any time in the future \cite{Bates2021}. Wadhams \cite{Wadhams2017} in his famous book 'farewell to ice', discussed this issue as unreasonable and he underlined the expectancy of an ice-free Arctic sooner.


Anderson \cite{anderson1961} summarized the general concerns about the Arctic sea-ice loss projection models as follows;

1- The sea-ice growth curve cannot be fitted by a single, simple power law such as results from the pure ice growth theories in commonly accepted literature \cite{stefan1891theorie, tamura1905mathematical}.

2- The relatively small scatter indicates that for a first approximation time and temperature may be compounded into a single parameter, namely, the exposure (degree-days or degree-hours of frost), and this is the most important parameter controlling the thickness of ice forming under a wide variety of conditions even on such a temperature-sensitive material. This commonly accepted procedure needs statistical justification since a proper theory of sea-ice growth requires knowledge of the actual thermal history, as well as such meteorological variables as humidity, wind velocity and radiation.

We can add further concerns which are related to the observation methods. Peng et al. \cite{peng2020} highlighted that the Albedo of the sea-ice area is calculated based on the sea-ice extent and the real Albedo might be much less than what is calculated, because of the melt ponds which are not taken into account. Due to resolution limitations, satellite based observations cannot show the open regions and the melt ponds \cite{Gao2021}. Because of these reasons, Markus Rex, Arctic mission chief explained that global warming may have reached an irreversible tipping point with these words \textit{"In the climate system, there are different tipping points, points that lead to sudden and irreversible changes. That could be triggered when global warming exceeds a certain threshold, and we have seen that the triggering of the tipping point that will lead irrevocably to the disappearance of the summer ice pack in the Arctic is imminent. Can we still save the year-round Arctic sea-ice? If we look at the completely melted ice at the North Pole in the summer of 2020, we can legitimately have doubts."} \cite{AFP2021}


Besides spatial resolution limitations, satellite observation data also has band resolution limitations which lead to more error in observation data. In the 2017 Carbon Brief article \cite{CarbonBrief2017} explained that the bands measured by the satellite instruments cannot easily provide the temperature of a specific layer of the atmosphere. Researchers have identified particular sets of bands that correspond to the temperature of the lower troposphere (TLT) spanning roughly 0 to 10 km, the middle troposphere (TMT) spanning around 0 to 20 km and the lower stratosphere (TLS) spanning 10 to 30 km. Unfortunately, these bands tend to overlap a bit. Both observations and models cause error. Benjamin et al. \cite{Benjamin2017} mentioned that, these forcings caused high uncertainties at the IPCC report models which are earlier than 2013 and the models will be updated in current modeling effort, called CMIP6, being done in preparation for the next Intergovernmental Panel on Climate Change report.

Shakhova et al. \cite{Shakhova2015} highlighted the danger of the accumulated methane (CH4) under the East Siberian Arctic Shelf and Arctic ocean which has been thawing more rapidly with the loss of the ice volume, may become a significant positive feedback to accelerate the global warming.

Last but not least, one of the most impactful self-reinforcing feedback parameters is the accumulated ocean heat content. Rising amounts of greenhouse gasses are preventing heat radiated from Earth’s surface from escaping into space as freely as it used to. Most of the excess atmospheric heat is passed back to the ocean. As a result, upper ocean heat content has increased significantly over the past few decades. More than $90\%$ of the warming that has happened on Earth over the past $50$ years has occurred in the ocean \cite{IPCC2013}. While the ocean itself is also absorbing the sunlight and probably keeping the radiated light within itself (not escaping) because of the ocean skin reflection impact \cite{Alley2000}

\textit{"The water’s heavier, and it has a higher specific heat, and both of those things give it a much bigger heat capacity. What this means for planet Earth is that excess energy might not make itself immediately obvious by strongly warming the atmosphere. Instead, that energy might hide in the ocean, in the form of warmer ocean temperatures."}
\cite{Scott2006}

Polyakov et al. \cite{Polyakov2020} and MacKinnon et al. \cite{MacKinnon2021} introduced the high heat content under the ocean which is not considered in the existing climate models. Yadav et al. \cite{Yadav2022} showed higher melting anomalies in the latest years, which are caused by not only the atmospheric heat and the albedo loss, but also because of the accumulated ocean heat content due to massive reduction of the ice volume which cannot offer the needed amount of the \textit{latent heat} budget.


\section{Methods}

We have started by plotting the total volume thickness of the sea-ice using the thickness values from Pan-Arctic Ice Ocean Modeling and Assimilation System (PIOMAS) data \cite{Zhang2003}. PIOMAS assimilates the daily sea-ice concentration and sea surface temperature satellite products. PIOMAS couples the Parallel Ocean Program (POP) with a 12-category thickness and enthalpy distribution (TED) model.  Uncertainties in PIOMAS ice thickness and snow depth result mainly from the model forcing (wind, thermal, and precipitation forcing etc.), physics, and parameterizations \cite{Schweiger2011}. The PIOMAS data covers the period 1979 to the present, providing estimates of some key ice and ocean variables including the ice thickness.

Melting ice uses the environmental heat for the transition process from the solid state to the liquid state. This state transition heat is not measurable, however it allows the ice to become a part of the ocean and it is known as \textit{latent heat}. When there is enough ice volume, the environmental heat is used within the \textit{latent heat} phase in the warm seasons. However, when there is not enough ice volume, the amount of the heat which is larger than the \textit{latent heat} phase requirements will stay in the environment and will cause further warming. 

\begin{figure}
\centering
\includegraphics[scale=0.29]{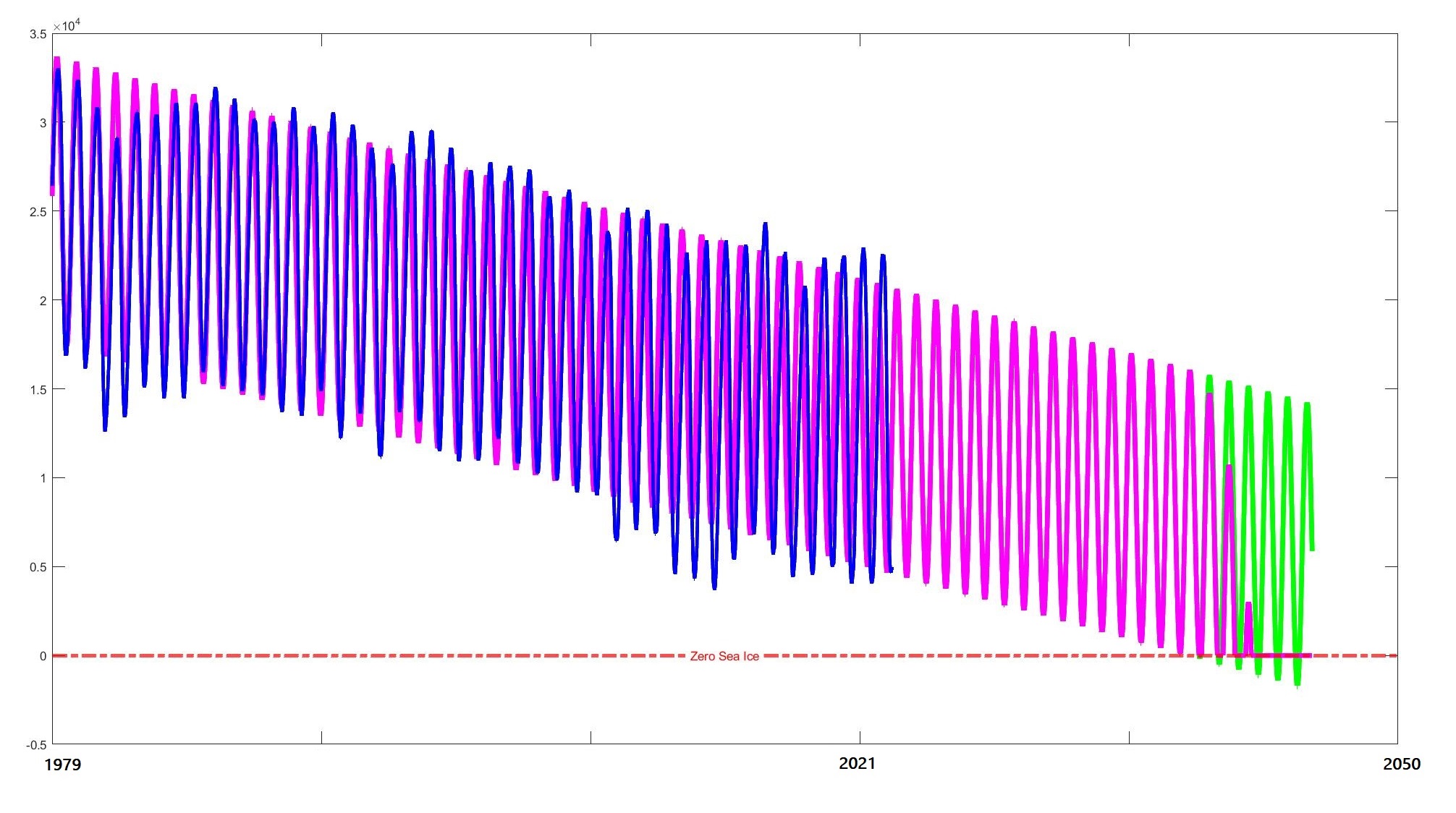}
\caption{Blue: PIOMAS data, Green: a sine wave which follows the same polynomial trend with the PIOMAS data, Magenta: the corrected sine wave after reducing the ice volume from the projection of each year, due to the increase of the ocean heat content. Horizontal axis corresponds to years and the vertical axis corresponds to the Arctic sea-ice volume.}
\label{projection}
\end{figure}

\begin{figure}
\centering
\includegraphics[scale=0.6]{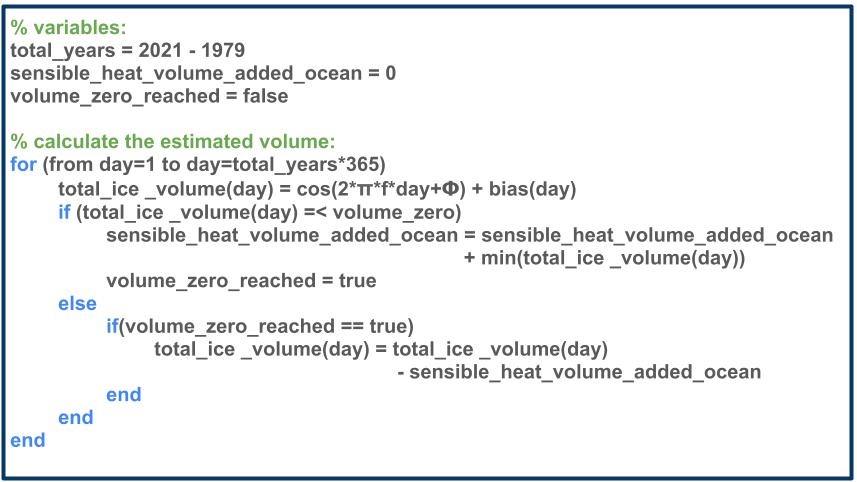}
\caption{Pseudocode for calculating the sea-ice projection values.}
\label{code}
\end{figure}

In Figure \ref{projection}, we have plotted the daily sea-ice volume measurements PIOMAS data from 1979 to the end of 2021 in dark blue color. With the pseudocode provided in Figure \ref{code}, we explained the algorithm that we used for creating a volume projection model. We calculated the $bias(.)$ function by finding the best fitting linear function to the mean of the PIOMAS data. We chose the linear bias function for simplicity, however possibilities of reaching to more accurate projection models with higher level polynomial bias functions should be investigated further. We have applied Fourier analysis to PIOMAS data to calculate the $f$ within the pseudocode and to generate a cosine wave which follows the same polynomial trend (green function in Figure \ref{projection}). $\phi$ value of the cosine function is used, in order to shift the cosine function towards to the yearly maximum values at its peak points. Finally, we have used the amount of the lost ice volume (under the \textit{Zero sea-ice} called red horizontal line) as a measurement which will cause further environmental heat (because of the lost \textit{latent heat} capacity) and allow decrease in the volume projections of the succeeding year. This projection is shown with the magenta color in Figure \ref{projection}. This projection shows the impact of the ice-free season in the sudden loss of Arctic before 2050, without showing any evidence of recovery. 

\section{Results and Discussion}

According to our research and experiments, we derive the following results.

\begin{figure}
\centering
\includegraphics[scale=0.5]{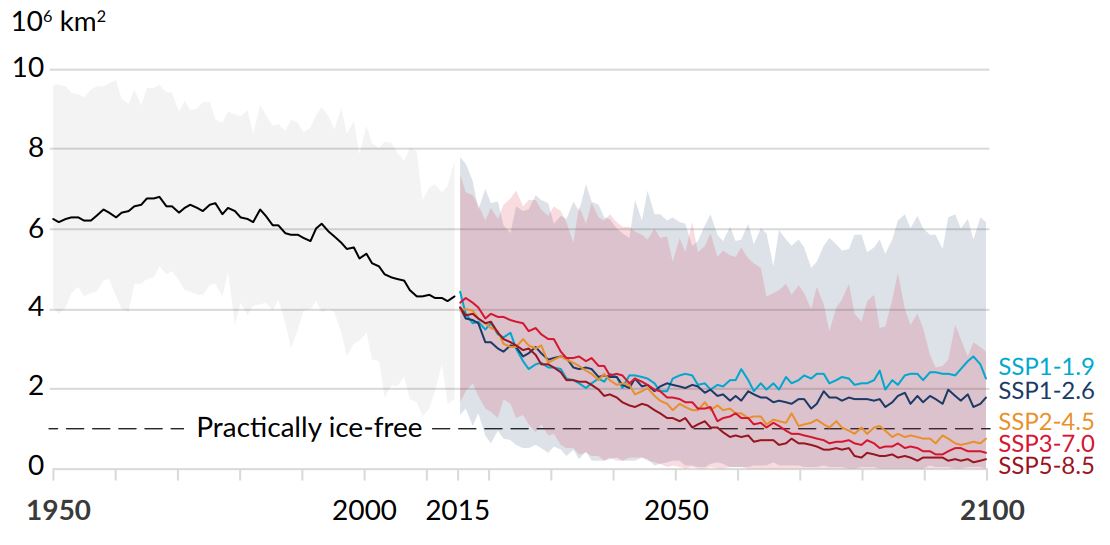}
\caption{September Arctic sea-ice coverage as predicted by different models (taken from IPCC AR6 full report) \cite{IPCC2021}}
\label{ipcc}
\end{figure}

\begin{itemize}
    \item Winter sea-ice in the Arctic region is likely to disappear in a few years (before 2050). In other words, \textbf{sea-ice loss is not reversible unlike how it was reported in IPCC AR6 and the earlier IPCC reports.}
    \item Loss of the winter sea-ice in the Arctic would cause high temperature increase and further environmental effects (such as significant slow down of the atmospheric and oceanic cycles) which is not represented in the commonly used climate models (since the ice-free Arctic and the related tipping point is not taken into consideration).
    \item Furthermore, the loss of the winter sea-ice would support CO2 and methane release from the ocean and the Siberian permafrost \cite{ZHAO2022105307}, which is not likely to be recovered by NetZero goals of the countries.
    \item All the evidence listed above shows that there is a tipping point in the Arctic. At this point, we differ from the IPCC. It is unlikely to not reach zero ice in the Arctic region as the IPCC AR6 report provided (Fig. \ref{ipcc}).
    \item When we considered the loss of the \textit{latent heat} capacity of the Arctic sea-ice volume within our projection model, the volume parameter showed a very sudden decrease. After the first ice-free year, in the next four years year-long ice-free region is predicted. Our analysis shows very high similarity to the previously occurred D-O events which showed very sudden disappearance of the sea-ice (in approximately five years according to the literature \cite{Bond1999}) after observing the first ice-free year. Although, the calculation of the years to reach the full year long ice-free region, depends on the level chosen for the horizontal red line in Fig. \ref{projection} (which represents the definition of the ice-free volume). In our study, we assumed that ice-free means that there is zero ice volume. However, in some studies "ice-free" definition accepts the appearance of some sea-ice to a certain extent. 
\end{itemize}

\section{Conclusions}

We discussed that the \textit{latent heat} capacity of the Arctic sea-ice reduces dramatically with the lost volume. This lost \textit{latent heat} capacity allows the environmental heat to be experienced as sensible heat and reduces even more ice volume. Thus, a self-reinforcing feedback loop is seen in the Arctic. Our PIOMAS data based analysis also validates that the loss of Arctic is unlikely to be reversible. Furthermore, we provided literature review on impact of the ice-free Arctic on global average temperature and climate systems. Our analysis and the literature review shows that there is a tipping point in the Arctic, which brings us to a different conclusion than the IPCC AR6 report which claims otherwise.







\bibliography{mybibfile}

\begin{thebibliography}{10}
\expandafter\ifx\csname url\endcsname\relax
  \def\url#1{\texttt{#1}}\fi
\expandafter\ifx\csname urlprefix\endcsname\relax\def\urlprefix{URL }\fi
\expandafter\ifx\csname href\endcsname\relax
  \def\href#1#2{#2} \def\path#1{#1}\fi

\bibitem{KHARE20211}
N.~Khare, R.~Khare, Chapter 1 - arctic—importance and physical structure
  (2021).
\newblock \href
  {http://dx.doi.org/https://doi.org/10.1016/B978-0-12-823735-9.00017-5}
  {\path{doi:https://doi.org/10.1016/B978-0-12-823735-9.00017-5}}.

\bibitem{Wang2020}
Z.~Wang, J.~Walsh, S.~Szymborski, M.~Peng,
  \href{https://journals.ametsoc.org/view/journals/clim/33/5/jcli-d-19-0528.1.xml}{Rapid
  arctic sea ice loss on the synoptic time scale and related atmospheric
  circulation anomalies}, Journal of Climate 33~(5) (2020) 1597 -- 1617.
\newblock \href {http://dx.doi.org/10.1175/JCLI-D-19-0528.1}
  {\path{doi:10.1175/JCLI-D-19-0528.1}}.
\newline\urlprefix\url{https://journals.ametsoc.org/view/journals/clim/33/5/jcli-d-19-0528.1.xml}

\bibitem{Yadav2022}
J.~Yadav, A.~Kumar, R.~Mohan, M.~Ravichandran,
  \href{https://doi.org/10.21203/rs.3.rs-1216059/v1}{Anomalous arctic sea ice
  melting linked to recent warming amplification}\href
  {http://dx.doi.org/10.21203/rs.3.rs-1216059/v1}
  {\path{doi:10.21203/rs.3.rs-1216059/v1}}.
\newline\urlprefix\url{https://doi.org/10.21203/rs.3.rs-1216059/v1}

\bibitem{IPCC2021}
V.~Masson-Delmotte, P.~Zhai, A.~Pirani, S.~Connors, C.~Pean, S.~Berger,
  N.~Caud, Y.~Chen, L.~Goldfarb, M.~Gomis, M.~Huang, K.~Leitzell, E.~Lonnoy,
  J.~Matthews, T.~Maycock, T.~Waterfield, O.~Yelekci, R.~Yu, B.~Z. (eds.),
  Ipcc, 2021: Climate change 2021: The physical science basis. contribution of
  working group i to the sixth assessment report of the intergovernmental panel
  on climate change, IPCC.

\bibitem{Schweiger2011}
A.~Schweiger, R.~Lindsay, J.~Zhang, M.~Steele, H.~Stern, R.~Kwok, Uncertainty
  in modeled arctic sea ice volume, Journal of Geophysical Research: Oceans
  116~(C8).

\bibitem{Alley2000}
R.~Alley, \href{https://doi.org/10.1073/pnas.97.4.1331}{Ice-core evidence of
  abrupt climate changes}, Proceedings of the National Academy of Sciences of
  the United States of America 97~(4) (2000) 1331–1334.
\newline\urlprefix\url{https://doi.org/10.1073/pnas.97.4.1331}

\bibitem{BrighamGrette18431}
J.~Brigham-Grette,
  \href{https://www.pnas.org/content/106/44/18431}{Contemporary arctic change:
  A paleoclimate d{\'e}j{\`a} vu?}, Proceedings of the National Academy of
  Sciences 106~(44) (2009) 18431--18432.
\newblock \href
  {http://arxiv.org/abs/https://www.pnas.org/content/106/44/18431.full.pdf}
  {\path{arXiv:https://www.pnas.org/content/106/44/18431.full.pdf}}, \href
  {http://dx.doi.org/10.1073/pnas.0910346106}
  {\path{doi:10.1073/pnas.0910346106}}.
\newline\urlprefix\url{https://www.pnas.org/content/106/44/18431}

\bibitem{NSIDC}
{NSIDC}, \href{http://nsidc.org/arcticseaicenews/2020/09/}{Arctic sea ice news
  and analysis, arctic sea ice decline stalls out at second lowest minimum}.
\newline\urlprefix\url{http://nsidc.org/arcticseaicenews/2020/09/}

\bibitem{Steffen8252}
W.~Steffen, J.~Rockstr{\"o}m, K.~Richardson, T.~M. Lenton, C.~Folke,
  D.~Liverman, C.~P. Summerhayes, A.~D. Barnosky, S.~E. Cornell, M.~Crucifix,
  J.~F. Donges, I.~Fetzer, S.~J. Lade, M.~Scheffer, R.~Winkelmann, H.~J.
  Schellnhuber, \href{https://www.pnas.org/content/115/33/8252}{Trajectories of
  the earth system in the anthropocene}, Proceedings of the National Academy of
  Sciences 115~(33) (2018) 8252--8259.
\newblock \href {http://dx.doi.org/10.1073/pnas.1810141115}
  {\path{doi:10.1073/pnas.1810141115}}.
\newline\urlprefix\url{https://www.pnas.org/content/115/33/8252}

\bibitem{Bond1999}
G.~Bond, W.~Showers, M.~Elliot, M.~Evans, R.~Lotti, I.~Hajdas, G.~Bonani,
  S.~Johnson, \href{https://doi.org/10.1029/GM112p0035}{The north atlantic's
  1-2 kyr climate rhythm: Relation to heinrich events, dansgaard/oeschger
  cycles and the little ice age}, In Mechanisms of Global Climate Change at
  Millennial Time Scales (eds P.U. Clark, R.S. Webb and L.D. Keigwin.
\newline\urlprefix\url{https://doi.org/10.1029/GM112p0035}

\bibitem{Rosenfeld2019}
D.~Rosenfeld, Y.~Zhu, M.~Wang, Y.~Zheng, T.~Goren, S.~Yu, Aerosol-driven
  droplet concentrations dominate coverage and water of oceanic low-level
  clouds, Science 363~(6427) (2019) eaav0566.
\newblock \href {http://dx.doi.org/10.1126/science.aav0566}
  {\path{doi:10.1126/science.aav0566}}.

\bibitem{KAUKER201613}
F.~Kauker, T.~Kaminski, M.~Karcher, M.~Dowdall, J.~Brown, A.~Hosseini,
  P.~Strand, Model analysis of worst place scenarios for nuclear accidents in
  the northern marine environment, Environmental Modelling and Software 77
  (2016) 13--18.
\newblock \href
  {http://dx.doi.org/https://doi.org/10.1016/j.envsoft.2015.11.021}
  {\path{doi:https://doi.org/10.1016/j.envsoft.2015.11.021}}.

\bibitem{sirmacek2022}
G.~McPherson, B.~Sirmacek, R.~Vinuesa, Environmental thresholds for
  mass-extinction events, Elsevier Results in Engineering.

\bibitem{Ruby2006}
L.~R. Leung, Y.-H. Kuo, J.~Tribbia,
  \href{https://journals.ametsoc.org/view/journals/bams/87/12/bams-87-12-1747.xml}{Research
  needs and directions of regional climate modeling using wrf and ccsm},
  Bulletin of the American Meteorological Society 87~(12) (2006) 1747 -- 1752.
\newblock \href {http://dx.doi.org/10.1175/BAMS-87-12-1747}
  {\path{doi:10.1175/BAMS-87-12-1747}}.
\newline\urlprefix\url{https://journals.ametsoc.org/view/journals/bams/87/12/bams-87-12-1747.xml}

\bibitem{PIDCOCK2014}
R.~Pidcock, Why aren’t climate models better at predicting arctic sea ice
  loss?,
  \url{https://www.carbonbrief.org/why-arent-climate-models-better-at-predicting-arctic-sea-ice-loss},
  [Online; accessed 27-11-2021] (2014).

\bibitem{Cohen2020}
J.~Cohen, X.~Zhang, e.~a. J.~Francis,
  \href{https://doi.org/10.1038/s41558-019-0662-y}{Divergent consensuses on
  arctic amplification influence on midlatitude severe winter weather}, Nature
  Climate Change 10 (2020) 20--29.
\newline\urlprefix\url{https://doi.org/10.1038/s41558-019-0662-y}

\bibitem{Bates2021}
J.~Bates,
  \href{https://greatwhitecon.info/wp-content/uploads/2021/12/Bates-Sea-Ice-Trends.pdf}{Polar
  sea ice and the climate catastrophe narrative}, The Global Warming Policy
  Foundation 28.
\newline\urlprefix\url{https://greatwhitecon.info/wp-content/uploads/2021/12/Bates-Sea-Ice-Trends.pdf}

\bibitem{Wadhams2017}
P.~Wadhams, A farewell to ice: A report from the arctic, Oxford University
  Press; Illustrated edition.

\bibitem{anderson1961}
D.~Anderson, Growth rate of sea ice, Journal of Glaciology 3~(30) (1961)
  1170--–1172.
\newblock \href {http://dx.doi.org/10.3189/S0022143000017676}
  {\path{doi:10.3189/S0022143000017676}}.

\bibitem{stefan1891theorie}
J.~Stefan, {\"U}ber die theorie der eisbildung, insbesondere {\"u}ber die
  eisbildung im polarmeere, Annalen der Physik 278~(2) (1891) 269--286.

\bibitem{tamura1905mathematical}
S.~T. Tamura, Mathematical theory of ice formation, Monthly Weather Review
  33~(2) (1905) 55--59.

\bibitem{peng2020}
H.-T. Peng, C.-Q. Ke, X.~Shen, L.~Mengmeng, Z.-D. Shao, Summer albedo
  variations in the arctic sea ice region from 1982 to 2015, International
  Journal of Climatology 40 (2020) 3008--3020.
\newblock \href {http://dx.doi.org/10.1002/joc.6379}
  {\path{doi:10.1002/joc.6379}}.

\bibitem{Gao2021}
P.~Gao, H.~Director, C.~Bitz, A.~Raftery, Probabilistic forecasts of arctic sea
  ice thickness, Journal of Agricultural, Biological and Environmental
  Statistics\href
  {http://dx.doi.org/https://doi.org/10.1007/s13253-021-00480-0}
  {\path{doi:https://doi.org/10.1007/s13253-021-00480-0}}.

\bibitem{AFP2021}
{AFP News Agency}, Arctic mission chief says global warming may have reached
  irreversible tipping point | afp, \url{https://youtu.be/9ZzqSOOkLu4},
  [Online; accessed 29-11-2021] (2021).

\bibitem{CarbonBrief2017}
{Carbon Brief}.

\bibitem{Benjamin2017}
B.~D. Santer, S.~Solomon, G.~Pallotta, C.~Mears, S.~Po-Chedley, Q.~Fu,
  F.~Wentz, C.-Z. Zou, J.~Painter, I.~Cvijanovic, C.~Bonfils,
  \href{https://journals.ametsoc.org/view/journals/clim/30/1/jcli-d-16-0333.1.xml}{Comparing
  tropospheric warming in climate models and satellite data}, Journal of
  Climate 30~(1) (2017) 373 -- 392.
\newblock \href {http://dx.doi.org/10.1175/JCLI-D-16-0333.1}
  {\path{doi:10.1175/JCLI-D-16-0333.1}}.
\newline\urlprefix\url{https://journals.ametsoc.org/view/journals/clim/30/1/jcli-d-16-0333.1.xml}

\bibitem{Shakhova2015}
N.~Shakhova, I.~Semiletov, V.~Sergienko, L.~Lobkovsky, V.~Yusupov, A.~Salyuk,
  A.~Salomatin, D.~Chernykh, D.~Kosmach, G.~Panteleev, D.~Nicolsky,
  V.~Samarkin, A.~C. S.~Joye, O.~Dudarev, A.~Meluzov, O.~Gustafsson, The east
  siberian arctic shelf: toward further assessment of permafrost-related
  methane fluxes and role of sea ice, Philosophical Transactions of the Royal
  Society A: Mathematical, Physical and Engineering Sciences 373~(2052).
\newblock \href {http://dx.doi.org/https://doi.org/10.1098/rsta.2014.0451}
  {\path{doi:https://doi.org/10.1098/rsta.2014.0451}}.

\bibitem{IPCC2013}
M.~Rhein, S.~Rintoul, S.~Aoki, E.~Campos, D.~Chambers, R.~Feely, S.~Gulev,
  G.~Johnson, S.~Josey, A.~Kostianoy, C.~Mauritzen, D.~Roemmich, L.~Talley,
  F.~Wang, Observations: Ocean. in: Climate change 2013: The physical science
  basis. contribution of working group i to the fifth assessment report of the
  intergovernmental panel on climate change [stocker, t.f., d. qin, g.-k.
  plattner, m. tignor, s.k. allen, j. boschung, a. nauels, y. xia, v. bex and
  p.m. midgley (eds.)], IPCC.

\bibitem{Scott2006}
M.~Scott, Earth's big heat budget, {NASA} earth observatory,
  \url{https://earthobservatory.nasa.gov/features/HeatBucket}, [Online;
  accessed 25-11-2021] (2006).

\bibitem{Polyakov2020}
I.~V. Polyakov, T.~P. Rippeth, I.~Fer, M.~B. Alkire, T.~M. Baumann, E.~C.
  Carmack, R.~Ingvaldsen, V.~V. Ivanov, M.~Janout, S.~Lind, L.~Padman, A.~V.
  Pnyushkov, R.~Rember,
  \href{https://journals.ametsoc.org/view/journals/clim/33/18/jcliD190976.xml}{Weakening
  of cold halocline layer exposes sea ice to oceanic heat in the eastern arctic
  ocean}, Journal of Climate 33~(18) (2020) 8107 -- 8123.
\newblock \href {http://dx.doi.org/10.1175/JCLI-D-19-0976.1}
  {\path{doi:10.1175/JCLI-D-19-0976.1}}.
\newline\urlprefix\url{https://journals.ametsoc.org/view/journals/clim/33/18/jcliD190976.xml}

\bibitem{MacKinnon2021}
J.~MacKinnon, J.~H. e.~a. H.L.~Simmons, A warm jet in a cold ocean, Nature
  Communications 12~(2418).
\newblock \href {http://dx.doi.org/https://doi.org/10.1038/s41467-021-22505-5}
  {\path{doi:https://doi.org/10.1038/s41467-021-22505-5}}.

\bibitem{Zhang2003}
J.~Zhang, D.~A. Rothrock, Modeling global sea ice with a thickness and enthalpy
  distribution model in generalized curvilinear coordinates, Monthly Weather
  Review 131~(5) (2003) 845 -- 861.

\bibitem{ZHAO2022105307}
S.~Zhao, W.~Cheng, Y.~Yuan, Z.~Fan, J.~Zhang, C.~Zhou, Global permafrost
  simulation and prediction from 2010 to 2100 under different climate
  scenarios, Environmental Modelling and Software 149 (2022) 105307.
\newblock \href
  {http://dx.doi.org/https://doi.org/10.1016/j.envsoft.2022.105307}
  {\path{doi:https://doi.org/10.1016/j.envsoft.2022.105307}}.

\end{thebibliography}

\end{document}